\documentclass[twocolumn,pra,aps,superscriptaddress,showpacs]{revtex4}
\usepackage{amsmath}
\usepackage{graphicx}

\begin{document}
\title{Exact solitons and manifold mixing dynamics in the spin-orbit coupled spinor condensates}
\author{Yong-Kai Liu}
\affiliation{Department of Physics, Beijing Normal University, Beijing 100875, China}
\author{Shi-Jie Yang\footnote{Corresponding author: yangshijie@tsinghua.org.cn}}
\affiliation{Department of Physics, Beijing Normal University, Beijing 100875, China}
\affiliation{State Key Laboratory of Theoretical Physics, Institute of Theoretical Physics, Chinese Academy of Sciences, Beijing 100190, China}
\begin{abstract}
We derive exact static as well as moving solitonic solutions to the one-dimensional spin-orbit coupled $F=1$ Bose-Einstein condensates. The static polar soliton is shown to be the ground state by the imaginary-time evolution method. It shows a helical modulation of the order parameter due to the spin-orbit coupling. In particular, the moving soliton exhibits a periodic oscillation among the particle numbers of the hyperfine states. We further explore the temporal evolution of the static polar soliton and find that the spin-polarization exhibits dynamical oscillations. This disappearance and reemergence of the ferromagnetic state indicates the mixing of the ferromagnetic and the antiferromagnetic manifolds.

\end{abstract}
\maketitle
\section{introduction}
In the past few years, the spin-orbit coupling (SOC) effects in multi-component Bose-Einstein condensates (BEC) have attracted great attentions\cite{Lin1,Lin2,Zhang,Lin3,Zhai2,Lan}. The experimental realization of synthetic magnetic fields in BECs open doors not only to provide an ideal pilot to model the SOC systems, but also bring out new physics which have not been considered before in the materials science, mainly due to the interplay between SOC and the unique properties of dilute atomic gases\cite{Barnett,Zhai}. One of the novel properties resulting from the nonlinearity in BECs is the existence of solitary-wave excitations\cite{Fialko}. For arbitrary spin-$F$ BECs, the order parameter has $2F+1$ complex components, producing a rich variety of spin textures\cite{Yuki3}.

Multi-component vector solitons of spin-1 BECs without SOC have been studied in Refs\cite{Ieda,Li,Szankowski1,Szankowski2}. Recently, some authors have studied the SOC effects on soliton structures in the two-component BECs\cite{Xu,Achilleos,Kartashov}. The spin-orbit coupling cooperates and/or competes with spin-dependent and spin-independent nonlinear interactions, giving rise to the bright solitons of exotic density profile and nontrivial dynamics as well. In this paper, we focus on the one-dimensional (1D) spin-1 BEC with SOC. We derive exact static as well as moving solitonic solutions to the coupled Gross-Pitaevskii equations (GPE). The static solution is proved to be the ground state by numerically evolving the GPEs along the imaginary-time. The textures originate from the helical modulation of the order parameter due to the SOC. We obtain the explicit relations of the characteristic parameters of the soliton with the various coupling strengthes. On the other hand, the moving solitonic solution displays a population oscillates among the hyperfine states due to the spin-orbit coupling.

In the temporal evolution of the static solitonic solution, we find that the spin-polarization oscillates dynamically which implies disappearance and reemergence of the ferromagnetic state even though the initial state is entirely constructed in the antiferromagnetic manifold. Our results manifest the manifold mixing physics in the spinor BEC concluded in Ref.{\cite{Oh}}.

The paper is organized as follows. In Sec. II we briefly describe the spin-1 BEC with SOC. In Sec.III we derive the static and moving solitonic solutions. In Sec.IV we explore the dynamical evolution of the static solution and reveal the manifold mixing. A discussion is included in Sec. V.

\section{Hamiltonian}
The mean-field order parameters of the spin-1 BEC are $\Psi(x,t)=(\psi_1, \psi_0, \psi_{-1})^T$. The scaled 1D Hamiltonian is $H=H_0+H_\textrm{int}$. $H_0=\int \Psi^{\dagger}
(\frac{1}{2}k^2-\alpha k\cdot \hat{\mathrm{f}}_y) \Psi dx$, where $\alpha$ characterizes the strength of the SOC and $\hat{\textbf{f}}$ are the spin matrices. The interaction term $H_\textrm{int}=\int dx (\frac{1}{2}c_0\rho^2+\frac{1}{2}c_2|\textbf{F}|^2)$, with $\rho=\sum_m |\psi_m(x)|^2$ the total density of the condensate and $\textbf{F}=\sum_m\psi_m^*\hat{\textbf{f}}_{mn}\psi_n$ the spin-polarization vector. The nonlinear coupling constants $c_0=(g_0+2g_2)/3$ and $c_2=(g_2-g_0)/3$, with $g_F$ relating to the $s$-wave scattering length of the total spin-$F$ channel as $g_F=4\pi\hbar^2a_F/M$\cite{Yuki3,Zhai2}.

The energy functional reads
\begin{eqnarray}
E=&&\int dx \{\frac{1}{2}|\partial_x \psi_1|^2+\frac{1}{2}|\partial_x \psi_0|^2+\frac{1}{2}|\partial_x \psi_{-1}|^2\nonumber\\
&&+\frac{\alpha}{\sqrt{2}}(\psi_1^*\partial_x \psi_0+\psi_0^*(-\partial_x \psi_1+\partial_x \psi_{-1})-\psi_{-1}^*\partial_x \psi_0)\nonumber\\
&&+\frac{c_0}{2}(|\psi_1|^2+|\psi_0|^2+|\psi_{-1}|^2)^2\nonumber\\
&&+\frac{c_2}{2}\times[(|\psi_1|^2-|\psi_{-1}|^2)^2+2|\psi_1^*\psi_0+\psi_0^*\psi_{-1}|^2]\}
\end{eqnarray}
The SOC term $i\alpha \mathrm{\hat f}_y \partial_x$ indicates that spin couples to the momentum in the $x$-direction. The ground state of a uniform system is ferromagnetic ($|\textbf{F}|=1$) for $c_2<0$ or polar ($|\textbf{F}|=0$) for $c_2>0$. For spinor BECs without the SOC, the Hamiltonian is invariant under the global $U(1)$ gauge transformation, the $SO(3)$ rotation in spin space, and time reversal $\mathcal{T}\equiv e^{-i\pi\mathrm{f}_y}\mathcal{K}$, $\mathcal{K}$ takes complex conjugation\cite{Kawaguchi1,Kawaguchi2}. For spinor BECs with SOC, the symmetry is largely reduced due to the requirement of simultaneous rotation in the spin space and the space\cite{Kawaguchi2}.

The stationary GPEs are written as
\begin{widetext}
\begin{eqnarray}
\mu\psi_1=[-\frac{1}{2}\bigtriangledown^2+(c_0+c_2)(\vert\psi_1\vert^2+\vert\psi_0\vert^2)+(c_0-c_2)
\vert\psi_{-1}\vert^2]\psi_1+c_2\psi_0^2\psi_{-1}^*+\frac{\alpha}{\sqrt{2}}\partial_x\psi_0,\nonumber\\
\mu\psi_0=[-\frac{1}{2}\bigtriangledown^2+(c_0+c_2)(\vert\psi_1\vert^2+\vert\psi_{-1}\vert^2)+c_0
\vert\psi_0\vert^2]\psi_0+2c_2\psi_0^*\psi_1\psi_{-1}-\frac{\alpha}{\sqrt{2}}(\partial_x\psi_1-\partial_x\psi_{-1}), \nonumber\\
\mu\psi_{-1}=[-\frac{1}{2}\bigtriangledown^2+(c_0+c_2)(\vert\psi_{-1}\vert^2+\vert\psi_0\vert^2)+(c_0-c_2)
\vert\psi_{1}\vert^2]\psi_{-1}+c_2\psi_0^2\psi_1^*-\frac{\alpha}{\sqrt{2}}\partial_x\psi_0,\label{stationary}
\end{eqnarray}
\end{widetext}
where $\mu$ is the chemical potential. The dimensionless wave functions $\psi$ satisfy the normalization condition $\int dx(|\psi_1|^2+|\psi_2|^2+|\psi_3|^2)=1$. In the next section we will construct an exact solution to the Eq.(\ref{stationary}) in the polar state $|\textbf{F}|\equiv 0$ manifold.

\section{solitonic solutions}
In order to derive an analytical solution to the GPE (\ref{stationary}), we focus on the polar state by taking the relation between the hyperfine states $\xi_{-1}=-\xi_1^*$ and $\xi^*=\xi_0$, where  $\psi=\sqrt{\rho(\textbf{x})}\xi(\textbf{x})$ and $\xi$ is a representative normalized spinor for the order parameter. This step simplifies the complex spin-spin couplings, which lead the 1D differential equations to be self-consistently solvable by making use of the properties of hyperbolic functions\cite{Yang}. We make a gauge transformation $\Psi=e^{i\alpha \mathrm{f}_y x}\tilde{\Psi}$ which yield to,
\begin{eqnarray}
\mu\tilde{\psi}_m&=&-\frac{1}{2}\bigtriangledown^2\tilde{\psi}_m+c_0\rho\tilde{\psi}_m+c_1\sum_n\textbf{F}\cdot\textbf{f}_{mn}\tilde{\psi}_n\nonumber\\
&-&\frac{1}{2}\alpha^2\sum_n(\mathrm{f} _y^2)_{mn}\tilde{\psi}_m.\label{stationary1}
\end{eqnarray}

Equation (\ref{stationary1}) has the usual bright solitonic solution for the attractive interaction ($c_0<0$), $(\tilde{\psi}_1,\tilde{\psi}_0,\tilde{\psi}_{-1})=A(0,\textrm{sech}(kx),0)$. For this form of solution the last term contributes a constant which is absorbed into the chemical potential. Hence we obtain a static solution as
\begin{equation}
\left(
          \begin{array}{c}
             \psi_1 \\
            \psi_{0}\\
             \psi_{-1}
          \end{array}
        \right)
=A\left(
          \begin{array}{c}
           -\frac{1}{\sqrt{2}}\sin(-\alpha x) \\
           \cos(-\alpha x)\\
           \frac{1}{\sqrt{2}}\sin(-\alpha x)
          \end{array}
       \right) \textrm{sech}(kx)\label{solution},
   \end{equation}
where $A$ and $k$ are real constants. Substituting the solution into the stationary Eq.(\ref{stationary}), we get $\mu=-\frac{1}{2}k^2=\frac{1}{2}c_0A^2$.
It is notable that the coefficients $A$ and $k$ are irrelevant to the SOC strength $\alpha$. It shows that the SOC affects the relative distribution of the density among the hyperfine states. The solution (\ref{solution}) is proved to be stable. We further demonstrate that the solution is exactly the ground state by taking a Gaussian-type initial state to the system and carrying out the imaginary time evolution. Figure 1 show that the analytical solution coincide with the ground state quite well.

\begin{figure}[h]
\includegraphics*[width=9cm]{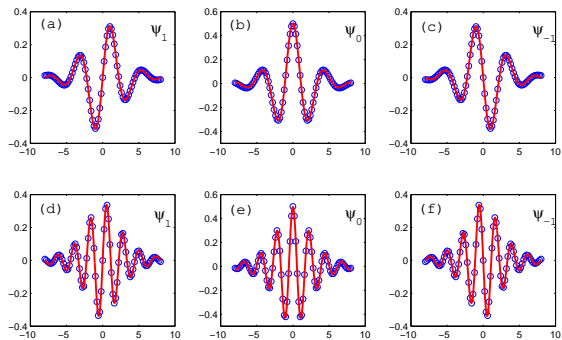}
\caption{(Color online) Static bright solitons for $c_0=-1$. Upper raw: $\alpha=\sqrt{2}$. Lower raw: $\alpha=2\sqrt{2}$. The real curves are numerical results by solving the GPEs with imaginary time evolution with a Gaussian initial wave package. The circles are from the analytical solution (\ref{solution}).}
\end{figure}

The bright soliton exhibits the $\sin(-\alpha x)$ or $\cos(-\alpha x)$ oscillating modulation under the influence of SOC. The oscillating frequency is directly related to the SOC strength. The nodes in these bright solitons are new feature in compare with the conventional BEC, which is also addressed in the spin-orbit coupled two-component BECs\cite{Xu}. The spin-1 matrix $\hat{\textbf{f}}$ have different properties as compared to the Pauli matrices $\hat{\sigma}$. For the real solution (\ref{solution}) with $|\textbf{F}|=0$, and $\xi_1=-\xi_{-1}$, the bright solitons shown in Fig.1 satisfy $\mathcal{T}\psi_m=\psi_m$, $\hat P\psi_m=(-1)^m\psi_m$, $\hat P$ is the parity operator and $\mathrm{\hat{f}}_y^2\psi=\psi$.

As to the polar state with $|\textbf{F}|\equiv 0$, the vector $\textbf{d}$ is introduced to describe the order parameter as $\xi=(\frac{-d_{x}+id_{y}}{\sqrt{2}},d_{z},\frac{d_{x}+id_{y}}{\sqrt{2}})^{T}$\cite{Yuki3}. From Eq.(\ref{solution}), we note that the $\textbf{d}$ is changed from a constant vector to a cycloidally varied vector with $d_y\equiv 0$ which corresponds to a planar spin. It carries a nonzero winding number and form a 1D skyrmion so far as the boundary condition is satisfied, as stated in Ref.{\cite{Kawakami}}.

The more general polar solitonic solution can be constructed as
\begin{equation}
\Psi=A\left(
          \begin{array}{c}
           \varepsilon\cos(\alpha x)+\sqrt{\frac{1}{2}+\sqrt{2}\varepsilon^2}\sin(\alpha x) \\
           -\sqrt{2} \varepsilon\sin(\alpha x)+\sqrt{1-2\varepsilon^2}\cos(\alpha x)\\
           -\varepsilon\cos(\alpha x)-\sqrt{\frac{1}{2}+\sqrt{2}\varepsilon^2}\sin(\alpha x)
          \end{array}
       \right) \textrm{sech}(kx),
   \end{equation}
where $\varepsilon\in[0,1/\sqrt{2}]$. We have numerically verified that all of these states are degenerate in energy and are energetically stable by imaginary-time evolution method.

\begin{figure}[b]
\includegraphics*[width=9cm]{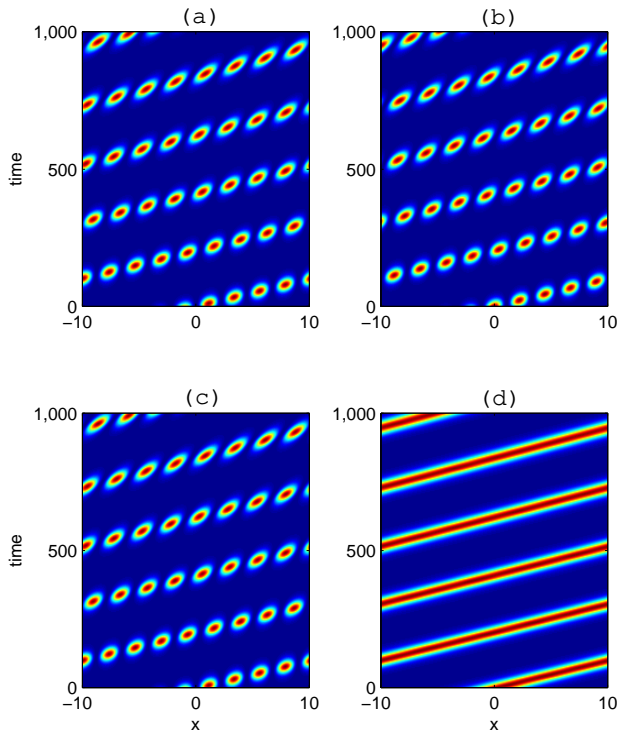}
\caption{(Color online) Real time evolution of the moving soliton for a initial wave package of solution (\ref{solution6}) at $t=0$. The parameters are $c_0=-1$, $c_2=1.5$, $\alpha=\sqrt{2}$. (a) $|\psi_1|^2$, (b) $|\psi_0|^2$, (c) $|\psi_{-1}|^2$, (d) $|\psi_1|^2+|\psi_0|^2+|\psi_{-1}|^2$.}
\end{figure}

Due to the absence of Galilean invariance, it is no longer a trivial task to give a moving soliton for a BEC with SOC\cite{Xu}. We still find a moving solitonic solution of the following form:
\begin{equation}
\Psi=A\left(
          \begin{array}{c}
           \frac{1}{\sqrt{2}}\sin(\alpha x) \\
           \cos(\alpha x)\\
           -\frac{1}{\sqrt{2}}\sin(\alpha x)
          \end{array}
       \right) \textrm{sech}(k(x-vt))e^{ivx}e^{\frac{i}{2}(k^2-v^2)t}\label{solution6},
   \end{equation}
where $v$ is the velocity of the soliton. Figure 2 show the real time dynamics of the moving solitonic solution (\ref{solution6}) with the velocity $v=0.1$, we observe that the density profiles of the hyperfine states exhibit periodic oscillations, due to the effects of spin-orbit couplings.

\section{Dynamics}
Now we study the dynamics of the static solitons (\ref{solution}) by adding a small perturbation and tracing the real time evolution of the GPEs. The results are displayed in Fig.3, where the interaction strength $c_0=-1$, $c_2=1$ and $\alpha=\sqrt{2}$. At first sight, the particle number in each hyperfine state is nearly unchanged and the density profile is almost stable. In fact, it oscillates with an small amplitude which indicates a small amount of ferromagnetic part appears in the condensate. For other parameters like $c_2=1.5$ and $c_2=0.5$, we have the same result. We mention that the result will be the same even we do not add the perturbations.

\begin{figure}[h]
\includegraphics*[width=9cm]{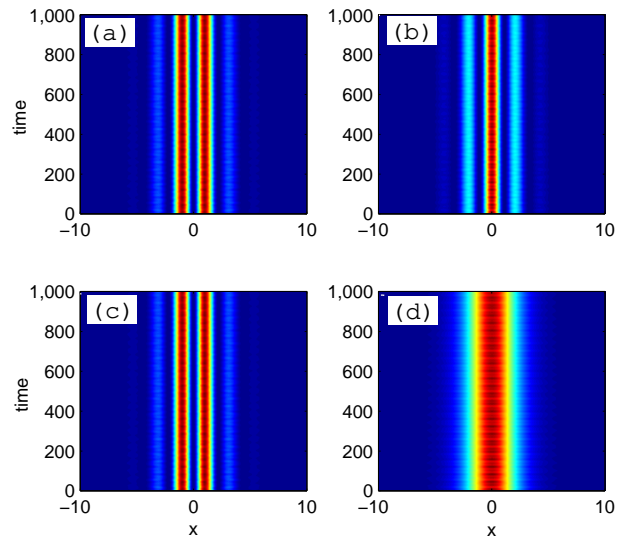}
\caption{(Color online) Real time evolution of the polar solution (\ref{solution}) by initially adding a perturbation. The parameters are $c_0=-1$, $c_2=1$, and $\alpha=\sqrt{2}$. (a) $|\psi_1|^2$, (b) $|\psi_0|^2$, (c) $|\psi_{-1}|^2$, (d) $|\psi_1|^2+|\psi_0|^2+|\psi_{-1}|^2$.}
\end{figure}

In the following, we will investigate the underlying physics of the oscillations in the dynamical evolution of the static soliton.
As we know, a dominant mechanism for destroying the knot (3D soliton) in spin-1 polar BEC is the spin current caused by the spatial dependence of the $\textbf{d}$ field\cite{Kawaguchi5}. The spin current induces the local magnetization according to \cite{Yuki3}
\begin{equation}
\frac{\partial\textbf{F}}{\partial t}
=\frac{\hbar}{M}\sum \textbf{d}\times \partial_i(\rho\partial_i\textbf{d})\label{polarization} ,
\end{equation}
which will destroy the polar state. The same mechanism applies in the present case. By substituting the wavefunction Eq.(\ref{solution}) into Eq.(\ref{polarization}), it is straightforward to calculate the time evolution of the local magnetization as $\frac{\partial F_y}{\partial t}|_{t=0}=2\alpha \rho k\textrm{tanh}(kx)$. Consequently, the spin-polarization will always evolve in time even the initial polar state is stable. This can also be verified by numerical simulations. Let the initial polar state (\ref{solution}) evolve for one short time $\Delta t$, we note that the expectation value of the spin $f_x=0$, $f_z=0$, and $f_y\neq0$. It implies that the temporal evolution cannot occur solely within the polar manifold for $t>0$.

\begin{figure}[tb]
\includegraphics*[width=9cm]{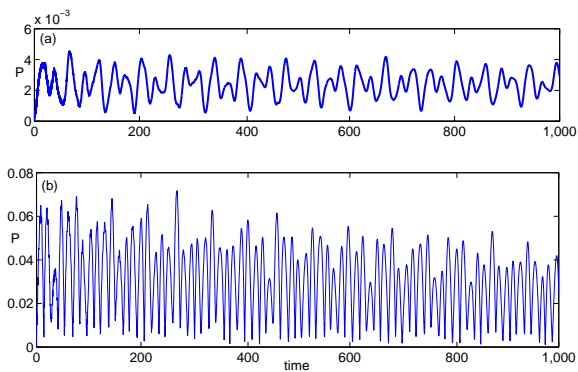}
\caption{(Color online) Evolution of the spin-polarization of (\ref{solution}) (a) without and (b) with initial perturbations, respectively. The parameters are $c_0=-1$, $c_2=1$ and $\alpha=\sqrt{2}$.}
\end{figure}

This phenomenon is related to the dynamical mixing of the polar ($\textbf{d}$) and ferromagnetic ($\textbf{F}$) manifolds. In order to give a qualitative analysis, we study the quantity $P=\int dx|\textbf{F}|$ to show the extent deviating from the polar state. $P=0$ indicates the state is completely the polar state while $P=1$ is a fully ferromagnetic state. From Fig.4, we observe that the value of $P$ becomes nonzero which oscillates with time. In Fig.4(a) without initial perturbation, the emergence and disappearance of the ferromagnetic part is very small, i.e., $P_{\textrm{max}}<0.005$. In Fig.4(b) with initial perturbation, a relatively large ferromagnetic part $P_{\textrm{max}}<0.08$ emerges. This situation can be understand as follows: Once the system starts to evolve, the initial polar state with the well-defined $\textbf{d}$-vector and inhomogeneous density will induce a nonzero polarization $\textbf{F}$, leading to the mixing dynamics of the polar and ferromagnetic states. The dynamical evolution of $\textbf{F}$ forms a competition with the $\textbf{d}$-vector, yielding a reciprocation of growth and reduction of the state that is close to the ideal polar state. Since the static solution is the exact and energetically stable, it does not show a large oscillation, in contrast to the 2D skyrmion case in antiferromagnetic spin-1 Bose-Einstein condensate which is energetically unstable\cite{Yip}.

A recent paper has studied the hydrodynamics theory of spin-1 BEC with the initial state defined solely within the ferromagnetic or antiferromagnetic manifold\cite{Oh}. It concludes that the system evolve into a mixture of the two manifolds, regardless of the sign and strength of the spin-dependent interaction. The analysis also applies to our case. In return, our results in the SOC spinor condensate manifest the theory of Ref.\cite{Oh}.

\section{Summary}
In summary, we have analytically presented static and moving solitonic solutions to the SOC spinor BECs. We find that the static solution is the ground state and the moving solution exhibits number oscillation among the hyperfine states. Due to the unique feature of the spin-1 BEC, the dynamics is distinct to the two-component BEC. The polar state will evolve into a state of mixed manifolds.
A feasible experiment mechanism is proposed by \cite{Lin3,Lan,Szankowski1}.

This work is supported by the NSF of China under grant No. 11374036 and the National Basic Research
Program of China (973 Program) under grant No. 2012CB821403.

\end{document}